\documentclass[3p,times,twocolumn]{elsarticle}
\usepackage{natbib}
\usepackage{graphicx}    
\usepackage{amssymb}    
\usepackage{amsmath}
\usepackage{url}
\usepackage{lineno}
\usepackage{hyperref}
\newcommand{\niffte}{NIFFTE}
\newcommand{\lansce}{LANSCE}
\newcommand{\ftpc}{fissionTPC}
\newcommand{\tof}{ToF}
\renewcommand{\u}[1]{\ensuremath{\mathrm{^{#1}}}U}
\newcommand{\pu}[1]{\ensuremath{\mathrm{^{#1}}}Pu} 

\newcommand{\hnel}{H(n,el)}

\newcommand{\En}{\ensuremath{E_n}}
\newcommand{\Ep}{\ensuremath{E_p}}
\newcommand{\Lp}{\ensuremath{L_p}}
\newcommand{\cossqth}{\ensuremath{{\cos^{2}\!\theta_p}}}
\newcommand{\csqt}{\cossqth}
\newcommand{\Ztpc}{\ensuremath{Z_\textsc{tpc}}} 
\newcommand{\xvar}{\ensuremath{x}}
\newcommand{\yvar}{\ensuremath{y}}
\newcommand{\zvar}{\ensuremath{z}}
\newcommand{\whichrho}{\ensuremath{r}} 
\newcommand{\Rhostart}{\ensuremath{\whichrho_\textsc{start}}}

\newcommand{\Rhotarget}{\Rhostart}
\newcommand{\Rhobeam}{\ensuremath{\whichrho_\textsc{beam}}}

\begin{document}
 

\begin{frontmatter} 
\title{$^1$H(n,el) as a Cross Section Reference in a White Source Neutron Beam With the fissionTPC}

\author[LLNL]{N.~I.~Walsh}\corref{cor1} 
\cortext[cor1]{Corresponding author} 
\ead{walsh25@llnl.gov}
%
\author[ACU]{Barker, J. T.}
\author[LLNL]{Bowden, N.~S.}
\author[ACU]{Brewster, K. J.}
\author[LLNL]{Casperson, R.~J.}
\author[LLNL]{Classen, T.}
\author[LANL]{Fotiadis, N.}   
\author[CSM]{Greife, U.}
\author[LANL]{Guardincerri, E.}
\author[LLNL]{Hagmann, C.}
\author[LLNL]{Heffner, M.} 
\author[CSM]{Hensle, D.}
\author[ACU]{Hicks, C.~R.}    
\author[LANL]{Higgins, D.}
\author[ACU]{Isenhower, L.~D.}
\author[CAL]{Kemnitz, A.} 
\author[ACU]{Kiesling, K. J.}
\author[OSU]{King, J.}
\author[CAL]{Klay, J.~L.}
\author[CSM]{Latta, J.} 
\author[OSU]{Loveland, W.}
\author[LLNL]{Magee, J.~A.}
\author[LLNL]{Mendenhall, M.~P.}
\author[LLNL]{Monterial, M.}  
\author[LANL]{Mosby, S.}
\author[CAL]{Oman, G.}
\author[LLNL]{Sangiorgio, S.}
\author[LLNL]{Seilhan, B.}
\author[LLNL]{Snyder, L.}
\author[ACU]{Towell, C.~L.}
\author[ACU]{Towell, R.~S.}
\author[ACU]{Towell, T.~R.}   
\author[LANL]{Schmitt, K.~T.}
\author[ACU]{Watson, S.}
\author[OSU]{Yao, L.}
\author[LLNL]{Younes, W.}
\author[]{\protect\\(The NIFFTE Collaboration)}
\address[LLNL]{Lawrence Livermore National Laboratory, Livermore, CA 94550, United States}
\address[LANL]{Los Alamos National Laboratory, Los Alamos, NM 87545, United States}
\address[CSM]{Colorado School of Mines, Golden, CO 80401, United States}
\address[ACU]{Abilene Christian University, Abilene, TX 79699, United States}
\address[CAL]{California Polytechnic State University, San Luis Obispo, CA 93407, United States}
\address[OSU]{Oregon State University, Corvallis, OR 97331, United States}


\date{\today}
   
\begin{abstract} 
  We provide a quantitative description of a method to measure neutron-induced fission cross sections in ratio to elastic hydrogen scattering in a white-source neutron beam with the fission Time Projection Chamber.  This detector has measured precision fission cross section ratios using actinide references such as \u{235}(n,f) and \u{238}(n,f).  However, by employing a more precise reference such as the \hnel{} cross section there is the potential to further reduce the evaluation uncertainties of the measured cross sections.  In principle the fissionTPC could provide a unique measurement by simultaneously measuring both fission fragments and proton recoils over a large solid angle.  We investigate one method with a hydrogenous gas target and with the neutron energy determined by the proton recoil kinematics.  This method enables the measurement to be performed in a white-source neutron beam and with the current configuration of the fissionTPC.  We show that while such a measurement is feasible in the energy range of $0.5$~MeV to $\sim$10~MeV, uncertainties on the proton detection efficiency and the neutron energy resolution do not allow us to preform a fission ratio measurement to the desired precision.  Utilizing either a direct measurement of the neutron time-of-flight for the recoil proton or a mono-energetic neutron source or some combination of both would provide a path to a sub-percent precision measurement.

\end{abstract}

\end{frontmatter}

\section{Introduction} 
\label{section:introduction}
The Neutron Induced Fission Fragment Tracking Experiment (\niffte{}) collaboration constructed the fission Time Projection Chamber (\ftpc) to measure fission cross section ratios of the major actinides (\u{235}, \u{238}, \pu{239}).  The aim of the experiment is to provide ratio measurements with  sub-percent uncertainties.  In the original design both actinide (n,f) and \hnel{} cross section references were considered~\cite{NIFFTEfissionTPC}.  To accommodate either option the detector is capable of detecting both fission fragments and proton recoils.  A description of a cross section measurement in ratio to an actinide reference can be found in Ref.~\cite{NIFFTEU8U5}. 

As the precision of ratio measurements are improved, the uncertainty on the cross section reference becomes an important factor.   The recent ENDF-B/VIII-0 evaluation lists nine neutron cross section standards~\cite{ENDF8}, with \hnel{} having the smallest uncertainty above 1\,MeV.  For example at $2\,$MeV the evaluated \hnel{} cross section uncertainty is $0.36\%$ compared to $1.3\%$ for \u{235}(n,f) and \u{238}(n,f)~\cite{CARLSON2018}.

While other experiments have measured a fission cross section relative to hydrogen elastic scattering, most relied on separate systems to detect fission fragments and protons (e.g.~Ref.~\cite{Nolte07}).  This potentially introduces a systematic uncertainty from uncontrolled differences in beam scattering between the two targets.  One experiment that made such a measurement in the same apparatus was performed in a mono-energetic beam~\cite{BartonUH}.  A back-to-back target design was used which significantly reduces beam scatter between the two targets. However, the proton measurement was limited in solid angle and separate detectors were used for fragments and protons.  

A ratio measurement with the \ftpc{} is unique as it can simultaneously detect both fission fragments and protons over a large solid angle with the same detector. %
In this work we examine one possible implementation of this measurement using a hydrogenous gas target and a determination of the neutron energy based on the proton recoil kinematics.  To perform this measurement over a broad neutron energy spectrum we examine the outcome using a beam source such as the 90L location at Los Alamos Neutron Science Center (\lansce{}) Weapons Neutron Research (WNR) facility \cite{lisowski06}.  We have made this choice of energy reconstruction and beam source as they do not require significant modification to the detector or signal processing compared to an actinide-to-actinide ratio measurement~\cite{NIFFTEU8U5}.

While this work is motivated in the context of a precise fission measurement, the method presented is applicable to neutron flux and spectrum measurements performed with a TPC. Specifically we describe in detail a charged-particle detection efficiency and neutron energy resolution which are relevant to, for example, measurements of a neutron beam flux~\cite{Maire2014}, directional neutron imaging~
\cite{LLNLnTPC,FU2018}, and cosmogenic neutron flux measurements~\cite{DCTPC}.

%
%
\section{The fissionTPC}
\label{section:detector}
\begin{figure}  
    \centering
    \includegraphics[width=\linewidth]{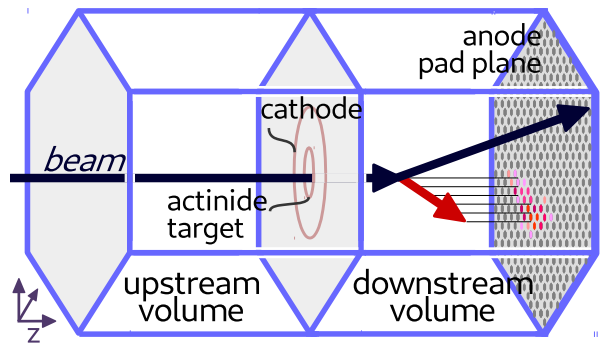}
    \caption{
      A simplified schematic of the \ftpc{} detector. The neutron beam impinges from the left  passing through the upstream volume, the actinide target, and the downstream volume. Both upstream and downstream volumes are 5.4\,cm in length and each instrumented with approximately 3000 conductive pads at the anode plane for read-out.  A charged particle track (represented by the red arrow) ionizes the gas and the charge is drifted towards the segmented anode.  The track is reconstructed using the pad's location and the relative time of arrival of the charge.
    } 
    \label{fig:TPCsketch}
\end{figure}

The \ftpc{} is a two-volume ionization chamber with a common cathode and two highly-segmented anode planes.  A negative bias is applied to the central cathode causing ionization electrons to drift towards the anode. The charge is amplified with a MICROMEGAS~\cite{GIOMATARIS} and read out from approximately 3000 conductive pads on each anode plane.  The target, typically an actinide deposited on aluminum or a thin carbon foil, is placed in the center of the cathode.  Depending on the backing thickness, one or both fission fragments induce a current on the cathode which is capacitively coupled to a current amplifier.  This provides the signal used to measure the incident neutron time-of-flight (\tof{}) in fission measurements.  A schematic of the detector with the relevant structures labeled is provided in Fig.~\ref{fig:TPCsketch}.  A more detailed description of the design is found in Ref.~\cite{NIFFTEfissionTPC}.
 
\begin{figure} 
  \centering  
  \includegraphics[width=1.0\linewidth]{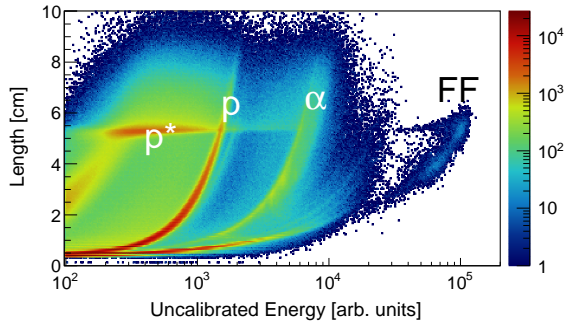}
  \caption{ 
    Data from the \ftpc{} taken at the \lansce-WNR neutron beam.  From left to right the labeled features are protons that do not stop in the detector (p*), contained protons (p), alphas ($\alpha$), and fission fragments (FF).  Particles traversing the entire drift volume are at least 5.4\,cm in length and produce a visible feature in the plot at that length.
  }
  \label{fig:data_len_vs_adc}  
\end{figure}

The \ftpc{} captures the following information on charged particle tracks: the vertex, direction, length, total charge, and ionization profile.  Two of the track's spatial dimensions ($\xvar, \yvar$) are reconstructed from the location of the anode pad.  The relative length along the drift axis ($\zvar$) is determined for all tracks by the time difference between the start and end of a track as measured on the anode.  The absolute position along the drift axis requires using the drift speed and the time difference between the cathode signal and anode signal.  In practice however, all fission fragments are assumed to have originated from the center cathode plane.  Most proton recoils, especially those further from the cathode, do not deposit enough energy to be detected on the cathode.  Therefore, this work assumes only the $\zvar$-length of a proton track is known.
Track energy and ionization profile are determined from the charge collected in the pads.  For this analysis, we reduce 2-dimensional ionization profile information to a maximum $dE/dx$ value (Bragg peak).  These track parameters (energy, length, and Bragg peak) are used to distinguish between types of particles.  An example of the distribution of length vs.\ energy for the different particle species in the \ftpc{} is shown in Fig.~\ref{fig:data_len_vs_adc}.

%
\section{Measurement Method}  
\label{section:method}

A cross section of reaction $x$ measured in ratio to reference $r$ is given by
\begin{equation}                   \label{eq:ratio}
  \frac{ \sigma_x }{ \sigma_r } = 
      \frac{C_x-B_x}{C_r-B_r} \cdot 
        \frac{N_r}{N_x} \cdot 
        \frac{\Phi_r}{\Phi_x} \cdot  
        \frac{\epsilon_r}{\epsilon_x} 
\end{equation}
where $C$ represents the number of measured reaction products, $B$ the number of background counts in that signal, $N$ the number of target atoms, $\Phi$ the beam fluence, and $\epsilon$ the detection efficiency. With the exception of the target atom number, each term is a function of neutron energy.  The fluence ratio in the \ftpc{} is very close to one, with the sub-percent correction computed via simulation.  The number of actinide target atoms is determined from a measurement of the alpha activity and scaled by half life data.  The remaining terms, including the number of hydrogen targets, efficiency, background, and neutron energy, depend on the specific experimental conditions.

Both a hydrogenous gas target and a solid target are candidates for a measurement with the \hnel{} reference.  While the quantitative result for the efficiency and background depends on the precise choice of the solid or gaseous target, the evaluation method itself is similar.  In a precision measurement with the \ftpc{}, the uniformity of the target is necessary to avoid systematic effects arising from beam non-uniformities.  A solid target like polystyrene can be spin coated to high uniformity on smooth silicon wafers.  The thickness, density, and uniformity can be measured with the required accuracy with a combination of atomic force microscopy, ellipsometry, and X-ray reflectometry.  However, complications of depositing an actinide an the same backing or having to remove the polystyrene from the backing (to avoid the large Si(n,p) background) without changing the uniformity are left to further studies.  On the other hand, hydrogenous gases like isobutane are used regularly in the detector and are intrinsically uniform.  The detector volume and gas properties such as pressure, temperature, and composition can be characterized to the required accuracy with commercially available equipment. 

In the \ftpc{} the cathode signal provides neutron \tof{} used for fission reactions.  A full width at half maximum (FWHM) timing resolution of 2\,ns or better provides sufficient resolution for precise fission cross section ratio measurements \cite{NIFFTEU8U5,TovHill_U}.  Extracting an accurate timing signal from protons is significantly more complicated and is not possible in the current \ftpc{} setup.  One of the complicating factors is the pile-up from multiple proton tracks, which is further compounded by the high rate of alpha tracks when measuring against a \pu{239} target.  In addition, the cathode detection efficiency decreases rapidly for proton tracks generated farther away from the cathode.  Therefore, we have chosen to investigate a kinematic method of reconstructing neutron energy using a gas target in the \ftpc{}.
In the kinematic reconstruction, the incident neutron energy ($\En$) is related to the scattered proton energy ($\Ep$) and polar angle with respect to the beam axis ($\theta_p$), and is given by  
\begin{equation} \label{eq:hnel} 
   \En   = \Ep / \csqt
\end{equation}
In the neutron energy range of interest ($<$10~MeV) the anisotropy is small so we assume the reaction is isotropic in the center-of-mass frame.

To investigate the feasibility of using a gas target and kinematic energy reconstruction, we have quantitatively evaluated the efficiency, backgrounds, and energy resolution.  These results are based on MCNP~\cite{MCNPX}, Geant4~\cite{geant4a, geant4b}, and the current NIFFTE analysis framework~\cite{Stave14}.  The MCNP simulation uses the neutron fluence based on the 90L station at LANSCE-WNR to generate the neutron-induced charged particles that enter the \ftpc{} detector volume. These charged particles are recorded and the vertices are used as the input to a Geant4 detector simulation.  The Geant4 simulation is interfaced with the NIFFTE analysis framework and together they account for the detector response, electronic read-out, and track reconstruction.  Inputs to the simulations have been chosen to closely approximate realistic detector conditions.  The charge amplification gain and gas properties were chosen to enable stable operation of the \ftpc{} when operated in high-energy neutron environment~\cite{NIFFTEgas}.
The simulations are performed using a gas mixture of neon and 5\% isobutane and total pressures in the range of 550 to 1500~Torr.  For each simulated pressure the electron diffusion and drift speed are estimated from MAGBOLTZ~\cite{MAGBOLTZ}.  The simulated gains and thresholds are chosen to match the observed length and energy distributions of data collected at 550 and 1000~Torr.
%
\section{Efficiency} 
\label{section:analysis:efficiency}  

The kinematic reconstruction of neutron energy requires the proton energy is fully deposited in the detector volume.  In a hydrogenous gas target, protons recoils are generated throughout the volume and with a range of energies and angles.  The likelihood of containment can be calculated because the direction and energy of a proton recoil is described exactly by two-body kinematics.  We describe two efficiencies, one based on the truth-level information of whether or not a track is contained and the second on a selection criteria based on the track-level information.

The probability a proton recoil is fully contained in the detector volume, the \emph{containment efficiency}, 
is a function of neutron energy, proton kinematics, and start position.  Summing over all possible proton kinematics and start positions, the efficiency is simplified to a function of only neutron energy. The containment efficiency can be computed numerically or by Monte Carlo and is determined by the reaction kinematics, detector geometry, and stopping power.  However in practice an analysis of data requires a selection gate on some track parameters to identify contained protons.  This \emph{selection efficiency} depends on not just the stopping power but also the detector performance, the tracking algorithms, and the selection criteria.   

\subsection{Containment Efficiency}

The containment efficiency is determined from the number of protons that stop in the detector relative to the total number of \hnel{} interactions in the volume. At each neutron energy, the probability a proton is contained is calculated summing over all interaction vertices and scattering energies (therefore all angles).  In the limit of low incident neutron energy, proton tracks are very short and almost always contained. In the other limit, when the proton length in the $\zvar$-direction exceeds the length of the detector, no vertex produces a contained proton.  Effectively the number of target atoms available to produce a contained proton scales according to the recoil kinematics as $(\Ztpc-\Lp\cos\theta_p)$, where $\Ztpc$ is the 5.4\,cm-length of the drift volume.  Assuming a beam radius ($\Rhobeam$) and a start vertex selection ($\Rhotarget$) not larger than the beam radius, the containment efficiency ($\epsilon$) is given by,
\begin{equation}             \label{eq:numer}
    \epsilon(\En)   = 
        \frac{ \left(\pi \Rhotarget^2\right) \cdot 
        \int \left(\Ztpc -\Lp\cdot\cos\theta_p\right)\cdot \mbox{d}\Ep  
             }{ \left(\pi \Rhobeam^2\right) \cdot \Ztpc}       
\end{equation}
A stopping power model is used to relate the proton track length ($\Lp$) and energy ($\Ep$).  Additionally, $\cos\theta_p$ is a function of $\Ep$ as given in Eq.~\ref{eq:hnel}. 
While this analytic form does not account for radial constraints of the detector or other more complicated geometries, these effects can be calculated with a toy model Monte Carlo.

\begin{figure}     
    \centering  
    \includegraphics[width=1.0\linewidth]{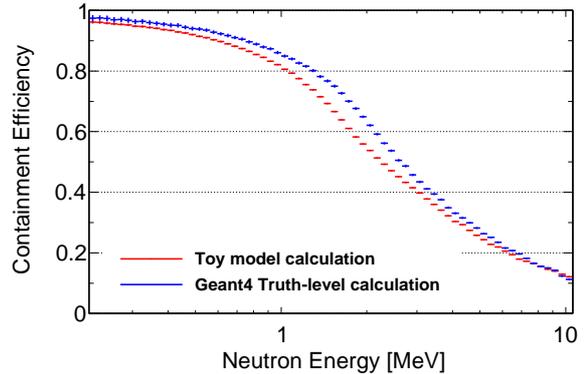} 
    \caption{  
      A toy Monte Carlo calculation of the containment efficiency for \hnel{} recoil protons to be fully contained in a detector with a geometry like the \ftpc{}. An efficiency calculated with a SRIM-based stopping power model (red) is compared to the Geant4 stopping power (blue).  The efficiency is largest at lowest neutron energies as only protons generated closest to the anode plane are long enough to not be contained. At the higher neutron energies, only protons from more glancing collision are contained. 
    }
    \label{fig:contain_effic}
\end{figure}

We have validated the numeric calculation of containment efficiency with a toy Monte Carlo.  In Fig.~\ref{fig:contain_effic} we show a comparison between a simple toy Monte Carlo based on the SRIM stopping power model~\cite{SRIM2010} and a Geant4 simulation. Although the Geant4 result also includes scattering in the gas and a more realistic beam profile, the largest difference between these efficiencies is from the stopping power.  
The difference in efficiencies shows that before a precision measurement can be performed, an accurate stopping power model should be identified and the uncertainties evaluated.

\subsection{Selection Efficiency}
\label{sec:particle_selection}

The selection efficiency is determined by the track-level identification of contained protons. It is sensitive to detector effects, the choice of tracking algorithms, and particle selections.  In this analysis we consider one realization of these tracking and selection choices.  

\begin{figure}     \centering               
  \includegraphics[width=1.0\linewidth]{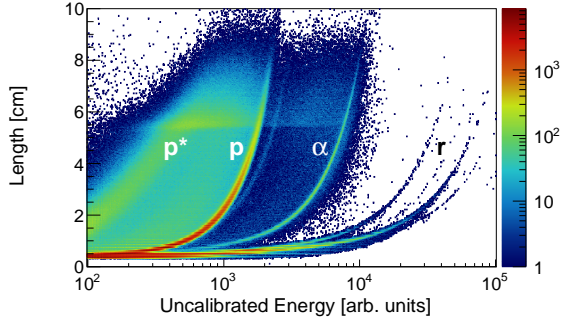}
  \caption{
    Simulated neutron-induced charged particles in the downstream volume of the \ftpc{} for the LANSCE-WNR neutron beam. The track length is plotted as a function of detected charge.  The main features of the plot from left to right are uncontained protons (p$^\star$), contained protons (p), alphas ($\alpha$), and ion recoils (r) dominated by carbon and neon.  The 5.4\,cm length of the drift volume accounts for the horizontal feature at that length.      
  }
  \label{fig:LenvADC_all}  
\end{figure}

\begin{figure}               
    \centering  
    \includegraphics[width=1.0\linewidth]{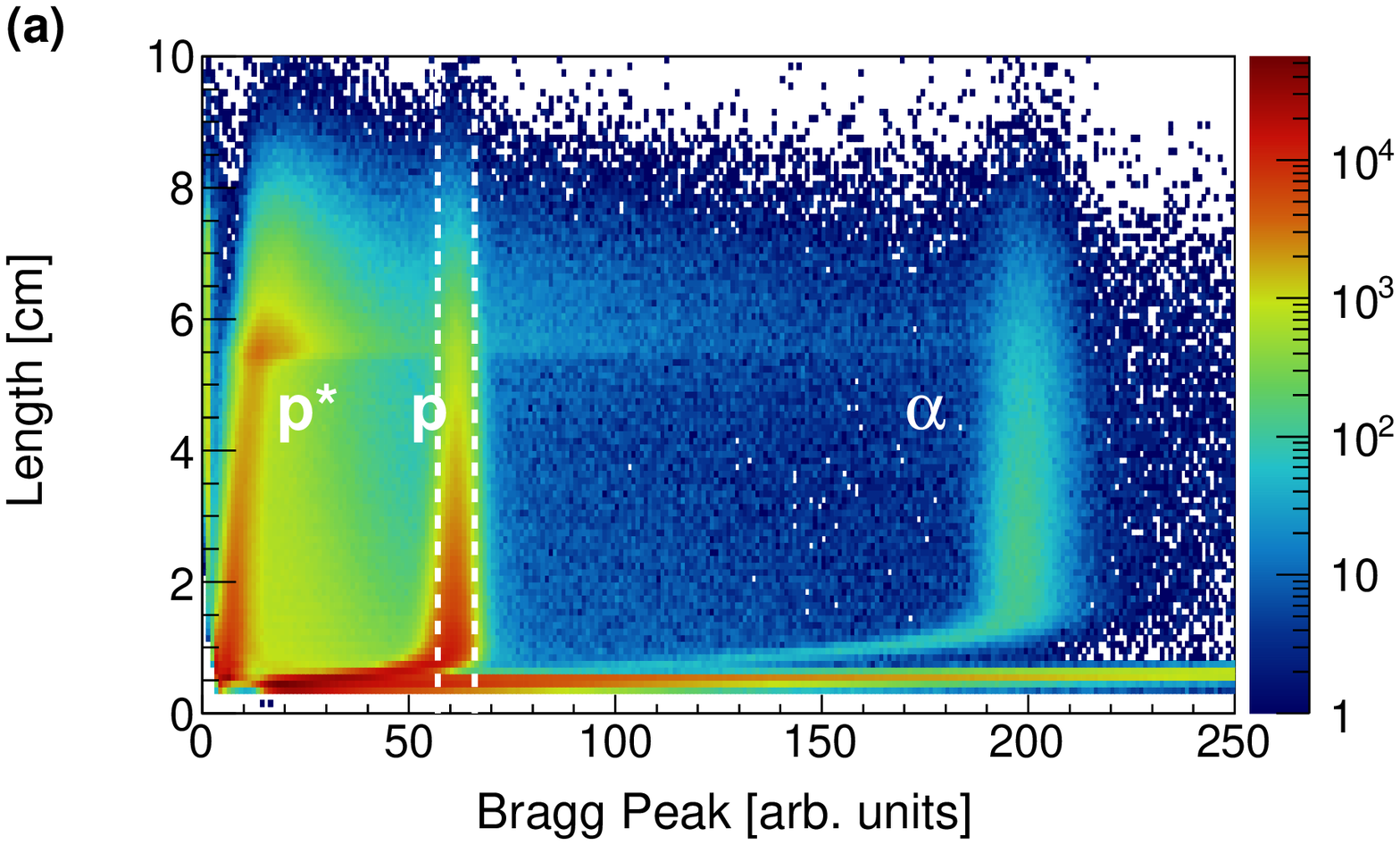}\\
    \includegraphics[width=1.0\linewidth]{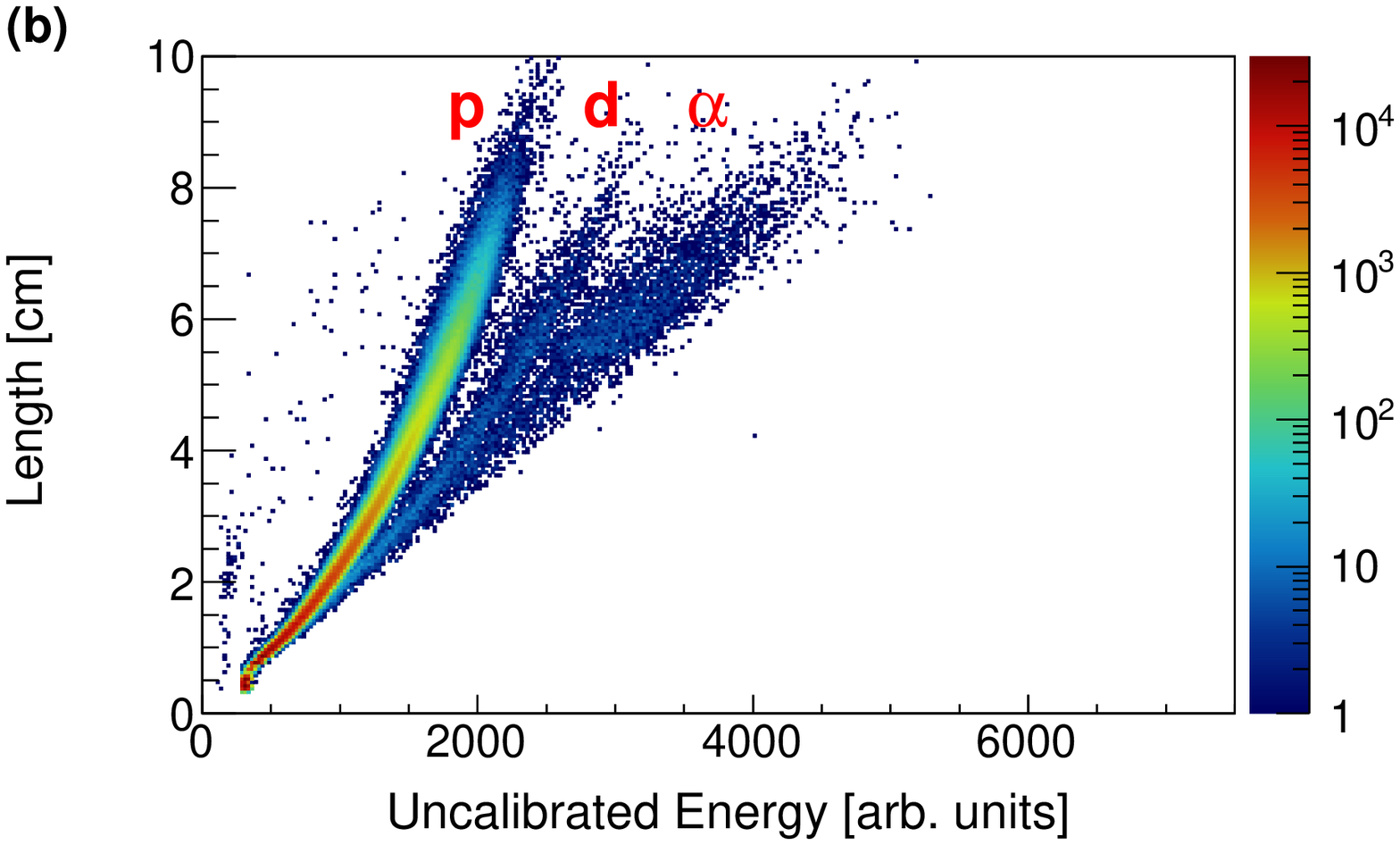}
    \caption{
      Distributions of light charged particles as simulated in the \ftpc{} showing (a) the Bragg peak (maximum $dE/dx$) and (b) the energy distribution after a selection on the Bragg peak for protons.  A selection of the proton Bragg peak between the dashed lines in (a) is used to generate the length-energy distribution in (b).  Additional cuts like a minimum length and a 2-dimensional cut on the proton length-energy band are used to improve the selection of contained protons.  The labels indicate features due to uncontained protons (p$^\star$), protons (p), deuterons (d), and alphas ($\alpha$). 
    }
    \label{fig:pid_cut}
\end{figure}

The 3-dimensional tracking and ionization profile information enables the separation of protons from other particles and the separation of contained protons from those not contained.  A distribution of length vs.\ energy in Fig.~\ref{fig:LenvADC_all} shows the neutron-induced protons, alphas, and ion recoils as simulated and reconstructed.  Contained protons are selected using a range of the maximum $dE/dx$ consistent with the proton Bragg peak.  This Bragg peak distribution and the resulting length vs.\ energy distribution after such a Bragg peak selection are shown in Fig.~\ref{fig:pid_cut}. Additional selections on length, polar angle, vertex, and energy further improve the identification of neutron-induced contained protons.
 

\begin{figure}     
    \centering  
    \includegraphics[width=\linewidth]{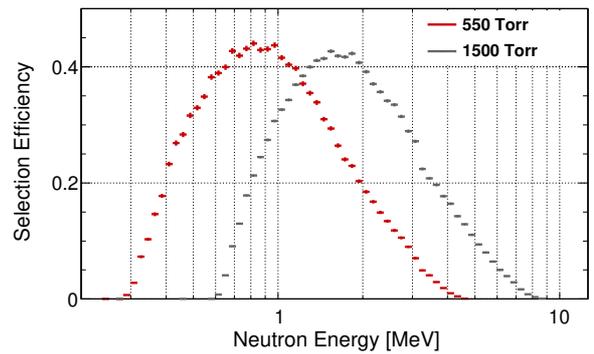}
    \caption{
      The selection efficiency computed from a simulation of the \ftpc{} with a neon-isobutane gas mixture at 550 and 1500~Torr.  This efficiency is based on a selection applied to the proton Bragg peak $dE/dx$ value to identify fully-contained protons from the \hnel{} reaction. This selection includes a minimum track length of 1\,cm and a minimum polar angle of $\theta\!<\!45^\circ$.  Error bars are statistical only.  
    }
    \label{fig:selection_effic}
\end{figure}

Based on the proton selection criteria applied to the simulation, we have computed the selection efficiency as a function of neutron energy.  As shown in Fig.~\ref{fig:selection_effic}, a broader range of neutron energies is accessible by operating the \ftpc{} at multiple pressures.  At the lower neutron energies the selection efficiency does not follow the containment efficiency because a minimum length cut eliminates recoils from low energy neutrons.  We apply a polar angle selection of $\theta\!<\!45^{\circ}$ which
limits the maximum neutron energy that can generate a contained proton.
%

\subsection{Uncertainty and Calibration}

The selection efficiency is subject to several sources of uncertainty related to the physics of stopping powers and electrons drifting in the gas.  Specifically it depends on the gas mixture, gas pressure, electron diffusion, drift speed, charge multiplication, and trigger threshold.  These are all included in the Geant4 modeling of the detector, but each would need to be calibrated for a precision measurement.

The models for electron diffusion, drift and stopping power can be calibrated to mono-energetic alpha decays of an actinide target.  A calibration to protons directly could be achieved in two ways.  The first is to use a source of mono-energetic neutrons such as from a DD generator. 
A second option is to use a neutron filter such as boron or carbon to effect a known distortion in the neutron spectrum (and therefore the proton spectrum).  This second method provides the ability to directly calibrate in a white source of neutrons.

The predicted selection efficiency may be validated with ratio measurements performed at multiple gas pressures and mixtures.  Each gas configuration has a different stopping power and therefore a different selection efficiency.  Each measurement is corrected for the efficiency and the different number of target atoms, with the flux normalization being obtained from an actinide (n,f) reaction and its cross section.

This validation relies on the assumption that the efficiency for detecting fission fragments does not change as a function of pressure. This is justified because the fission fragment source is localized along the z-axis and the primary driver for fission fragment efficiency is target thickness and neutron kinematics~\cite{NIFFTEU8U5}.

%
\section{Backgrounds}
\label{section:analysis:backgrounds}  
In this analysis tracks that pass the selection criteria that are not protons from \hnel{} reactions in the gas, including particles mis-identified as protons, are considered to be backgrounds. 
The dominant backgrounds arise from other neutron induced reactions with a proton in the final state.  Most such backgrounds are threshold reactions like (n,p) which have neutron energy thresholds of around $5$ to $10$~MeV.  These reactions produce track angles and energies that do not preserve the incident neutron energy information.  Without a time-of-flight to verify the kinematic reconstruction, these inelastic reactions are indistinguishable from elastic proton recoils.

\begin{figure}[hbt] 
  \centering  
  \includegraphics[width=1.0\linewidth]{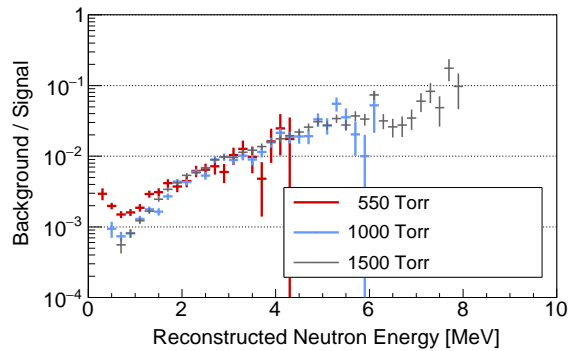}
  \caption{ 
    Simulated background rates relative to the \hnel{} reaction assuming a kinematic neutron energy reconstruction.  These results are based on the \ftpc{} with a neon-isobutane mixture operated in the LANSCE-WNR neutron beam.  Error bars are statistical only. The main backgrounds are (n,p) reactions caused by neutrons of higher energies than as reconstructed assuming an elastic collision.  Below 3\,MeV the background is less than 1\% of the signal.  At higher energies, corrections of up to 10\% are required. 
  }
  \label{fig:bkgd:three}  
\end{figure}

Beam-induced background protons originate from the detector vessel, target backing, anode planes, and the non-hydrogenous gas components.  The upstream side of the \ftpc{} is useful as a veto of tracks that pass through part of the upstream volume, through the cathode, and stop in the downstream volume.  These tracks are identified and removed if they are coincident and co-linear.  Another potential background source is back-scattered protons from the downstream anode plane. These are rejected based on their direction.  The remaining background sources are protons created in the central portion of the cathode (the target backing) and the downstream gas components.  Reducing the target-backing mass greatly impacts the background rate.  In this work a thin $100\,\mu$g/cm$^2$ carbon foil was chosen.  After applying these selections, the resulting background rates (Fig.~\ref{fig:bkgd:three}) relative to \hnel{} are expected to be less than $1\%$ below reconstructed neutron energies of $3\,$MeV. Corrections approaching 10\% are required up to 8~MeV.  The dominant background is the Ne(n,p) reaction.

%
\section{Neutron Energy Resolution} %
\label{section:analysis:resolution}  

In the kinematic reconstruction of the incident neutron energy, the energy resolution is determined by the combined proton energy and angular resolutions.  These resolutions ultimately depend on the specific experimental parameters such as electron drift speed, electron diffusion, gain, and thresholds. We evaluate these resolutions with the Geant4 simulation using a combination of estimated and measured parameters for the \ftpc{}.

While the simulation provides information about the detector effects, we have not explicitly evaluated the effects of calibration uncertainties on the proton energy and angle.  In the simulation the proton energy is determined with a linear scaling of the collected charge. In data this relation is determined using mono-energetic alphas of known energies such as those emitted from an actinide target.  The polar angle calibration is directly related to that of the drift velocity.  The drift velocity is set to a value that reconstructs the polar angle distribution of spontaneous alpha decay as an isotropic distribution.  The uncertainty is determined from a combination of statistics and variations from the fit range used to evaluate the polar angle isotropy.  In a previous measurement with a \u{235} alpha source, the drift velocity was determined with an uncertainty of 0.3\%. 

\subsection{Proton Energy Resolution} 
\label{subsection:resolution:proton}

In the \ftpc{} the charged particle energy resolution is impacted by variations in the charge-to-energy gain in each anode pad due to variations in the preamplifiers and the MICROMEGAS structure. Some of this variation is reduced by calibrating the pad gains for each run. Using  data from mono-energetic alphas, the charge voxel from each pad is first re-binned relative to the alpha track charge cloud axis.  Each pad is then compared to the volume-averaged distribution of each bin to provide an estimate of the gain correction.  
Pad gain variations of 5\% to 10\% are typical in the data while the variations were found to be stable to less than 1\%.  The stability allows for a reliable correction to be applied.  For reference, at a pressure of 550\,Torr the energy resolution of $4.4\,$MeV alphas after this calibration is $1.7\%$. 

\begin{figure*}    
  \centering
  \includegraphics[width=0.33\linewidth]{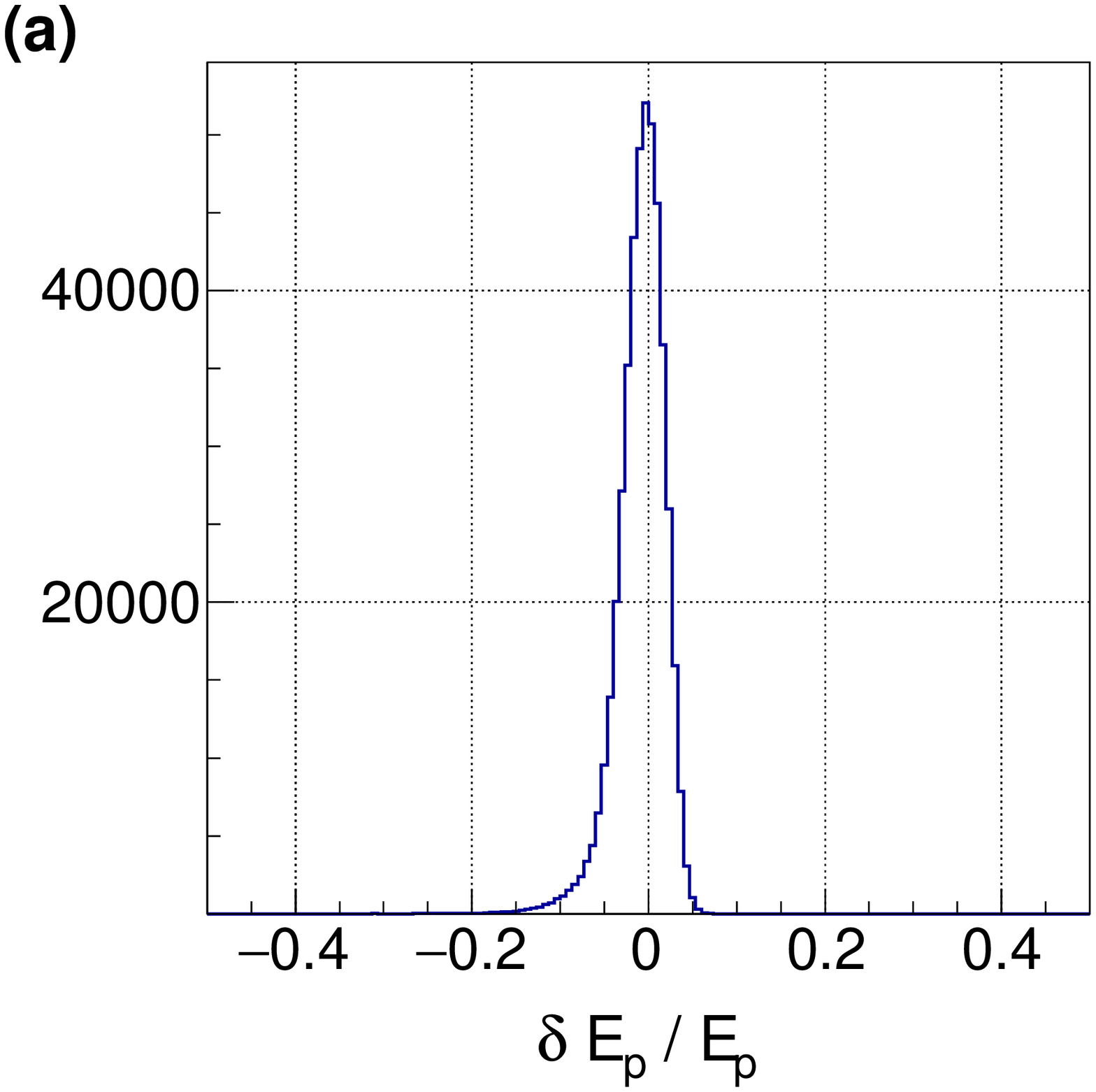}%
  \includegraphics[width=0.33\linewidth]{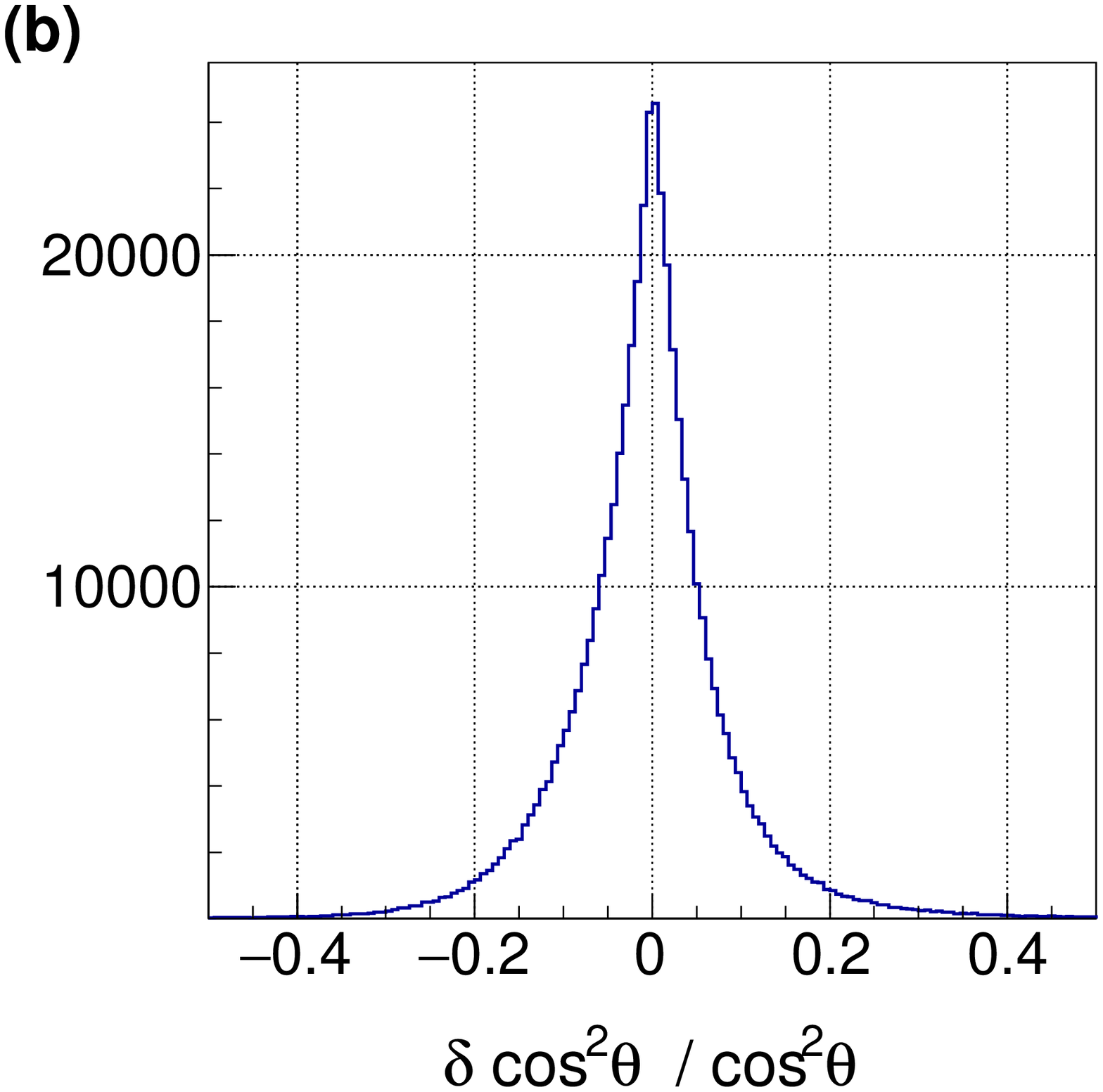}%
  \includegraphics[width=0.33\linewidth]{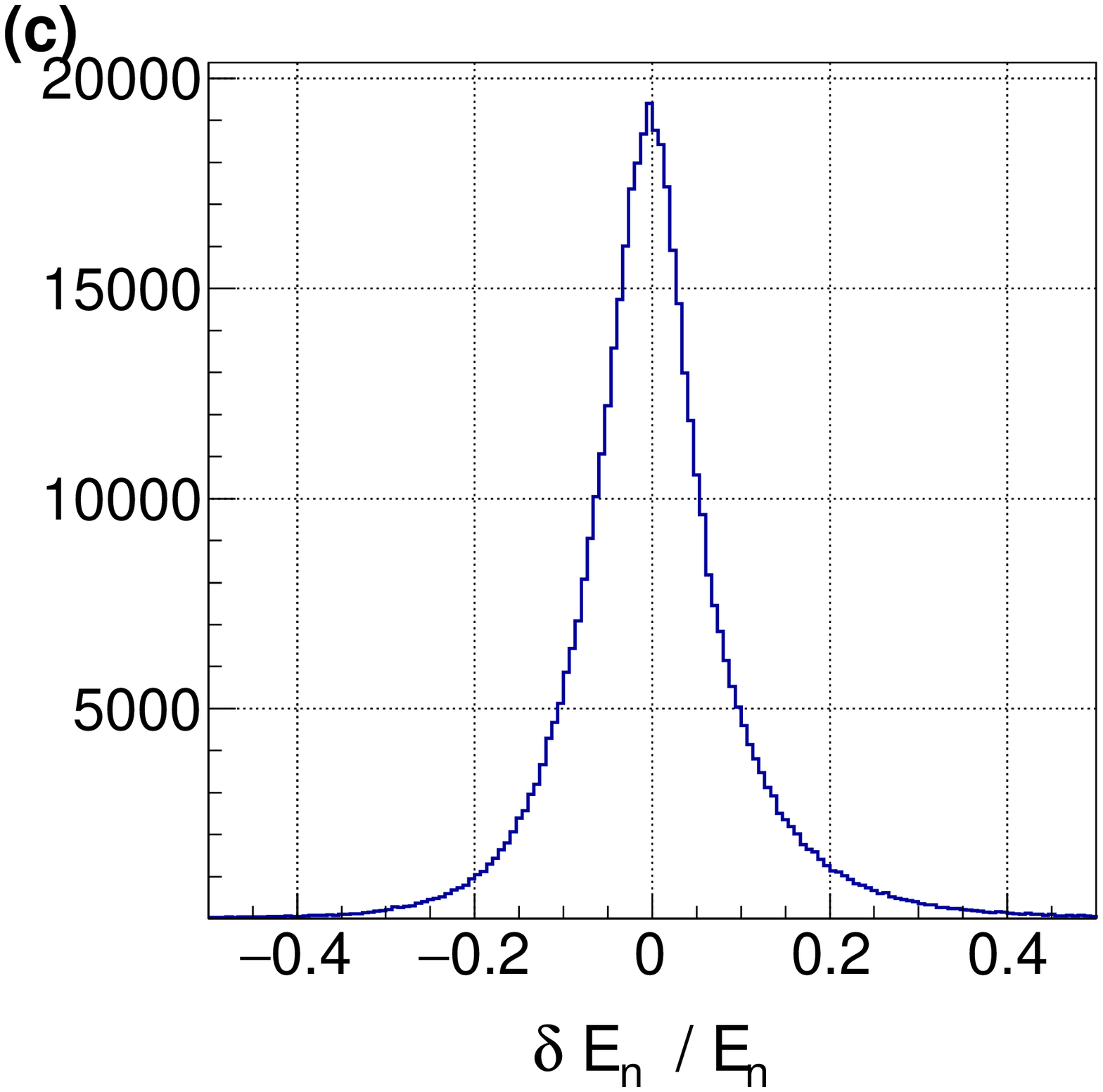}
  \caption{  
     Resolution on the reconstruction of (a) proton energy, (b) cosine of the polar angle squared, and (c) neutron energy generated from a simulation of neon-isobutane gas at 1000 Torr in the \ftpc{}.  Contained protons are identified by selections that include a minimum length ($>\!1$~cm), polar angle ($\theta\!<\!45^\circ$), and a Bragg peak selection.  The FWHM of the $\Ep$ resolution is 5\%, the $\cos^{2}\theta$ FWHM is 8\%, and the $\En$ FWHM is 12\%.  A polar angle selection of $\theta\!<\!20^\circ$ improves the angular and neutron energy resolution by nearly a factor of two.
  }
  \label{fig:reso_3}
\end{figure*}

Although in the simulation the pad thresholds and gains are uniformly applied, other detector effects are included and provide a reasonable estimate of the resolution.  The same selection of contained protons as in the previous sections is used to determine the proton energy resolution (Fig.~\ref{fig:reso_3}(a)). The 
FWHM of the distribution ranges from $\sim$8\% at a gas pressure of 550 Torr to $~\sim$4\% at 1500 Torr.  

The distribution is nearly symmetric except for a small tail where the reconstructed energy is less than expected.
We find these tail events are due primarily to at least one of three possible reasons. 
One reason is the proton is nearly contained but after depositing energy corresponding to its Bragg peak, the remaining energy is not deposited in the gas volume. 
A second reason is charge is lost because of diffusion.  While the charge cloud drifts towards the anode, it spreads out and at the edges the charge falls below the trigger threshold of the pad. The amount of energy lost increases with the amount of diffusion and therefore also with drift distance.  A third scenario which affects both the length and energy of the track, occurs at the beginning of the track where the stopping power is lowest and the charge deposited falls below the pad threshold. 
Some of these effects can be minimized by operating at pressures and drift fields that reduce the diffusion or by increasing the gain of the MICROMEGAS.

\subsection{Angular Resolution}       
\label{subsection:resolution:angular}

The angular resolution is computed by comparing the polar angle from the reconstructed track to the initial direction as determined from the \hnel{} kinematics.  This resolution is given in term of $\csqt$ as this scales with the neutron energy.  Unlike proton energy resolution, the angular resolution does not depend strongly on gas pressure. For example, at 1000 Torr and with a $\theta\!<\!45^\circ$ selection, the angular resolution, shown in Fig.~\ref{fig:reso_3}(b), has a FWHM of 8\%. 

The intrinsic few-degree scattering of protons stopping in a gas is the dominant contributor to this resolution. The impact of this is more significant at larger $\theta$ due to the cosine function.  The angular resolution improves to 5\% FWHM with a selection criteria of $\theta\!<\!30^\circ$ and to 3.5\% FWHM with $\theta\!<\!20^\circ$.
This cut improves the angular resolution but at the expense of a reduced selection efficiency and narrower accessible range of neutron energies.
Additionally, we identify tracking biases that arise due to the asymmetric value of electron diffusion parallel and perpendicular to the drift field. This effect can be measured and corrected for in the tracking algorithm.  Using an \emph{ad hoc} correction to adjust the $\cos\theta$ bias removes the skew in the resolution distribution, however this correction does not change the FWHM of the distribution.

\subsection{Neutron Energy Resolution}
\label{subsection:resolution:energy}

\begin{figure}
  \centering
  \includegraphics[width=1.0\linewidth]{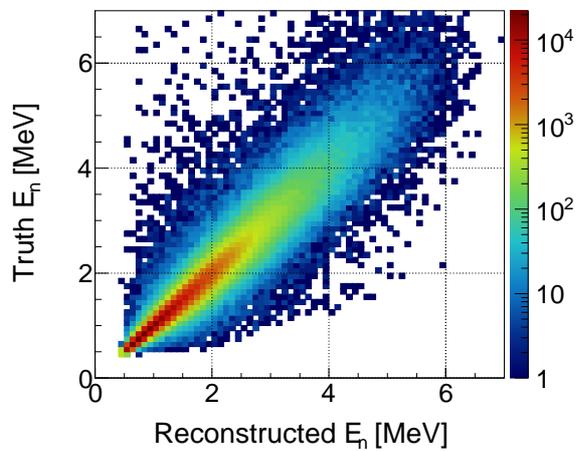}
  \caption{  
     Distribution showing the spread of the truth neutron energy vs.\ the reconstructed neutron energy using the kinematic method in a simulation of the \ftpc{}.
  }
  \label{fig:en_true_reco}
\end{figure}

\begin{figure} 
  \centering
  \includegraphics[width=\linewidth]{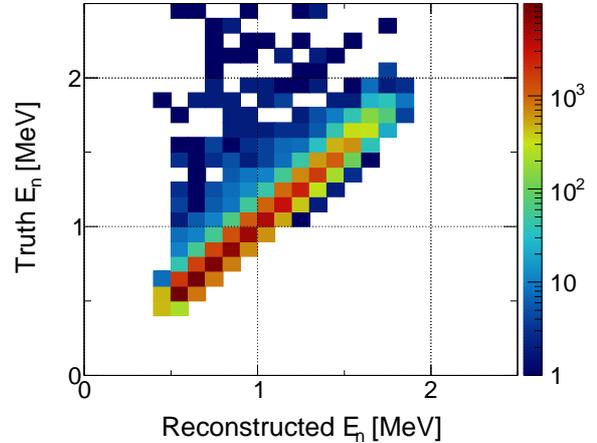}
  \caption{  
     Distribution of truth and reconstructed neutron energy similar to Fig.~\ref{fig:en_true_reco} but with a narrower selection of polar angles ($\theta\!<\!20^\circ$).  The few events in the tail of the distribution (truth energy greater than reconstructed energy) are from background (n,p) reactions.
  }
  \label{fig:en_true_reco_fwd}
\end{figure}

We compute the reconstructed neutron energy resolution with respect to the truth incident neutron energy. 
The resolution for the \ftpc{} detector is shown in Fig.~\ref{fig:reso_3}(c).
The FWHM of the neutron energy resolution varies from 12\% at 1500~Torr to 16\% at 550~Torr.  A selection of forward polar angles ($\theta\!<\!20^{\circ}$) improves the resolution to 7\%.  A distribution of the truth neutron energy versus the reconstructed energy in Fig.~\ref{fig:en_true_reco} displays the impact this resolution has on reconstructing the correct energy.
Similarly Fig.~\ref{fig:en_true_reco_fwd} shows the energy resolution after applying a forward angle selection cut of $\theta\!<\!20^{\circ}$.
Even with the coarse binning shown in the figures and a forward angle selection the effect of the resolution is substantial.  Events not along the diagonal of these plots would be events placed in the wrong energy bin.

Compared to the \tof{} method in the \ftpc{} with a 2\,ns FWHM timing resolution, the $\sim$10\% energy resolution from this method is, for example at 2~MeV, 10 times worse. 
Ultimately the impact from the energy resolution on a cross section ratio depends on the bin width and the slope and structure of the cross section ratio.  A measurement with only the kinematic method of energy reconstruction precludes using the \hnel{} cross section as a reference in a precision (n,f) measurement.  However, in other cross section measurements or neutron imaging experiments this resolution may be sufficient.

%
%
\section{Discussion and Conclusion}
\label{section:conclusion}
This work provides a quantitative assessment of a method to apply the precisely known \hnel{} cross section as a reference in a neutron-induced fission cross section ratio measurement.  We evaluate the efficiency, background, and energy resolution as they are all critical to the ratio measurement.  The decision to use a gaseous hydrogenous target and the kinematic energy reconstruction is motivated by the desire to operate the \ftpc{} with minimal detector development while at a white-source neutron beam.  The method presented is also relevant for neutron imaging experiments and neutron flux measurements performed with other TPCs.

We show that without knowing the absolute $\zvar$-position of the proton, we are able to compute a proton selection efficiency as a function of neutron energy.  
From our simulations, we show the ionization profile can be used to identify fully contained protons and based on a selection of this ionization profile we have calculated a selection efficiency.  The estimated background contribution assuming a thin-carbon backing and a hydrogenous gas target has a minimal effect below 3~MeV, with small corrections needed at higher energies.  The energy resolution has been evaluated and several systematic effects were identified.  Ultimately the energy resolution of this method is limited by the intrinsic few-degree scattering of protons stopping in the gas.

Although this measurement is feasible, it would not provide a sufficiently precise reference for a fission cross section ratio.  The precision is limited by the neutron energy resolution and a reliance on simulation for background and efficiency corrections.  Developing a robust method to extract neutron \tof{} from the proton recoil would greatly improve the neutron energy resolution.  A fast \tof{} signal would also eliminate the low energy background as a direct measurement of the incident neutron energy would make them distinguishable from signal protons where the kinematic method cannot.  Furthermore, a \tof{} measurement would remove the requirement that the proton be fully contained which will significantly reduce the complexity of the efficiency correction.

A measurement at a mono-energetic neutron facility is also an option towards a precision measurement.  A mono-energetic beam would eliminate the high-energy (n,p) backgrounds.  This would then allow for measurements to be made with a thick silicon target backing on which a solid hydrogenous target could be mounted.  A solid target at a fixed location in $\zvar$ rather than gaseous target also simplifies the efficiency correction. 
%
%
\section{Acknowledgments}
The neutron beam for this work was provided by LANSCE, which is funded by the U.S. Department of Energy and operated by Los Alamos National Security, LLC, under contract DE-AC52-06NA25396. This work performed under the auspices of the U.S Department of Energy by Lawrence Livermore National Laboratory under contract DE-AC52-07NA27344. This material is based upon work supported by the U.S. Department of Energy, National Nuclear Security Administration, Stewardship Science Academic Alliances Program, under Award Number DE-NA0002921.  

\bibliographystyle{elsarticle-num} 
\bibliography{biblio}
\end{document}